\documentstyle[aps,epsfig,prl]{revtex}
\begin{document}
\twocolumn[\hsize\textwidth\columnwidth\hsize\csname@twocolumnfalse\endcsname

\title{
Reentrant layering in rare gas adsorption: preroughening or premelting?
}
\author{
Franck Celestini$^{1}$,
Daniele Passerone$^{2,3}$,
Furio Ercolessi$^{2,3}$, and
Erio Tosatti$^{2,3,4}$
}
\address{
$^{\rm (1)}$ Laboratoire MATOP associ\'e au CNRS, Universit\'e 
d'Aix-Marseille III,
Marseille, France}
\address{
$^{\rm (2)}$ International School for Advanced Studies (SISSA-ISAS),
I-34014 Trieste, Italy}
\address{
$^{\rm (3)}$ Istituto Nazionale di Fisica della Materia (INFM), Italy}
\address{
$^{\rm (4)}$ International Centre for Theoretical Physics (ICTP),
I-34014 Trieste, Italy}

\maketitle

\begin{abstract}

The reentrant layering transitions found in rare gas adsorption on 
solid substrates have conflictually been explained either in terms of 
preroughening (PR), or of top layer melting-solidification phenomena.
We obtain adsorption isotherms of Lennard-Jones particles on an attractive 
substrate by off lattice Grand Canonical Monte Carlo (GCMC) simulation, and
reproduce reentrant layering. Microscopic analysis, including
layer-by-layer occupancies, surface diffusion and pair correlations,
confirms the switch of the top surface layer from solid to quasi-liquid
across the transition temperature. At the same time, layer occupancy is found
at each jump to switch from close to full to close to half, indicating a
disordered flat (DOF) surface and establishing preroughening as the
underlying mechanism. Our results suggest that top layer 
melting is essential in triggering preroughening, which thus represents
the threshold transition to surface melting in rare gas solids.

\end{abstract}
\vspace{0.5cm}
]

\narrowtext
Rare gas solid surfaces and films provide an important testing 
ground for a variety of surface phase transitions. Surface melting 
\cite{SM}, roughening \cite{vanbejeren}, and more recently preroughening (PR) 
\cite{dennijs} have been identified or at least claimed at the free rare gas
solid-vapor interface.
Layering transitions of thin rare gas films
on smooth substrates have given rise to a wide literature \cite{younbook}.
The discovery of reentrant layering (RL) -- the unexpected disappearance and subsequent
reappearance (well below the roughening temperature) of layering steps in adsorption isotherms on smooth substrates \cite{younhess,day} -- has led to a debate \cite{comment}. One possible explanation is PR,
 a phase transition which takes a surface from a low-temperature 
``ordered flat'' state, with essentially full surface coverage ($T<T_{PR}$), to
a high temperature 
``disordered flat'' (DOF) state, with half coverage, and a network of meandering 
steps ($T>T_{PR}$). 
Layering would disappear at PR, but re-enter in the DOF state \cite{dennijs,weichman}. The competing explanation is based on the
possibility of a melting-solidification-melting sequence in 
the top surface layer, similar to that seen for increasing temperature in
canonical molecular dynamics simulations \cite{phillips}.
In this picture, RL would result directly from a layer-promotion-driven 
melting of the top 
surface layer, and the subsequent advance of a solid-liquid interface \cite{comment}.
Both approaches appear to capture some important physics, but both
also have problems. 
The non-atomistic statistical mechanics lattice models provide, in presence of an attractive substrate potential, an overall adsorption phase diagram with zig-zag
lines of heat capacity peaks (whose behavior has
been called ``zippering'' \cite{zippering}) which
are centered at $T_{PR}$ and strikingly resemble 
experimental observations \cite{day}.
Because they contain PR, the models can naturally explain why the coverage jump across RL should 
be about half a monolayer, as seen in ellipsometry \cite{younhess} and in X-ray measurements \cite {rieutord}.
However, they fail to account for continuous atom dynamics, in particular melting, and
it remains unclear how bad the total neglect of these aspect might be at these relatively high temperatures. In Ar (111), RL takes place near $69 K$,
not too far from melting at $T_m=84 K$. By contrast, the atomistic canonical simulation approach does not suffer from that problem, and can 
describe quite well all the surface degrees of freedom, including thermal evolution of
each surface layer from 
solid to liquid.
It finds, realistically, that top-layer surface melting seems to be setting on 
precisely near the RL temperature.
However, it does not explain the half layer coverage jump across RL.  
A crucial underlying difficulty of this approach lies in the fixed
particle number -- a difficulty which the lattice models, being naturally
grand canonical, do not encounter.
In this situation Grand Canonical Monte Carlo (GCMC) atomistic simulation should be the method of choice,
applied since long ago \cite{rowley} to describe adsorption,
albeit of a single monolayer. Recently, we demonstrated
how a free rare gas (111) surface 
can be realistically simulated by GCMC with a Lennard-Jones potential,
and found indications that PR is indeed incipient 
at $0.8\,T_{m}$ \cite{franck1}. 
That work however remained incomplete, because a full equilibrium 
stabilization of the grand canonical surface proved 
to be increasingly hard with increasing temperature, and failed 
above $ 0.8 T_m$, where a value of $\mu$ that would cause neither decrease nor increase of
the total particle number could no longer be found.

In this Letter we present results of a fully equilibrated 
realistic GCMC simulation of multilayer
rare gas adsorption on a flat attractive substrate. In this case, 
the substrate potential naturally provides the necessary
stabilization for the system. We obtain realistic adsorption isotherms, 
whose main features compare directly with experiment. Reentrant layering is 
recovered, and layer occupancies confirm its association with 
a DOF surface and thus with PR.
At the same time however, surface  diffusion and pair correlations
show that while the virtually full monolayer below $T_{\rm PR}$ is solid, 
with only a gas of adatoms and vacancies, the half-full monolayer found 
above $T_{\rm PR}$ is made of a 2D liquid islands (even if in a strong periodic potential).
A new picture emerges, where the fractional monolayer melting, besides 
opening the way to surface melting, is also 
a key element favoring the preroughening of these surfaces.
\begin{figure}
\centerline{\epsfig{figure=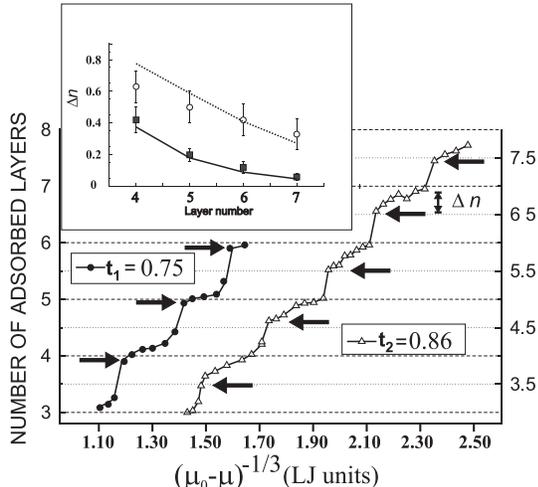,clip=,width=7cm}}
\caption{  Calculated adsorption isotherms for $t_1= 0.75$ ($0.53 \epsilon$) 
and $t_2= 0.86$ ($0.61\epsilon$), respectively below and above reentrant
layering ($t= 0.83$). Arrows indicate that at $t_1$ the layering steps
lead to roughly integer coverage, while at $t_2$  they lead
to roughly half-integer coverage. 
The abscissa for isotherm $t_2$ is shifted by $0.4$.
Inset: amplitude of the
 plateau breadths between two jumps in the $t_2$ isotherm. 
Circles: this simulation. Squares: extracted from 
Ar/graphite data\protect\cite{younhess}.
Full line: model fit to experimental breadths, as described in the text,
with $c/k = 1200$.
Dashed line: prediction of the same model, for a 10-fold enhanced
substrate attraction $\gamma$.
}
\end{figure}

We simulate adsorption by classical GCMC, implementing small displacement
moves (m),
creations (c), and destructions (d) with relative probabilities 
$\alpha^{(m)}=1-2\alpha$ and $\alpha^{(c)}=\alpha^{(d)}=\alpha$. 
Small moves apply to all particles, whereas creation/destruction 
is restricted to a fixed surface region, about four layers wide, since their
acceptance in the fourth layer of this region is already negligible on the
entire MC run.
In standard bulk GCMC the fastest convergence to the Markov chain 
is for $\alpha=1/3$ \cite {norman}. For our surface geometry and 
our potential, the optimal value of 
$\alpha$ is found to be small, of order $10^{-3}$ (the precise value 
depending on the outer layer population relative to the total), as needed to
allow for a more effective equilibration after each creation/destruction move.
Creation and destruction acceptance probabilities were
checked explicitly to satisfy the detailed balance.
We simulated adsorption of atoms interacting via the (12,6) Lennard-Jones
potential truncated at $2.5 \sigma$. 
The bulk fcc triple point temperature $T_m$ of this model is $\sim 0.7 \epsilon$ \cite {xjc} 
 (note that pressure dependence is negligible, {\it i.e.} $P_m/T_m (dT_m/dP) \simeq 2 \times 10^{-4}$ for Ar),
and we will from now on switch notation to a reduced temperature $ t = T/T_m$.
The substrate was taken to be flat and unstructured. Periodic 
boundary conditions  were assumed along the
$x$ and $y$ directions, with a reflecting wall along $z$, placed way above the surface. Interactions between atoms and substrate were also
of the Lennard-Jones form, giving rise to a laterally
invariant (3,9) potential  $V(z)=A (B/z^9 - C/z^3)$, with 
$A=40 \pi/3$, $B=1/15$ and $C=1/2$, the latter $\simeq 10$ times 
larger than the true Ar/graphite value, so as to
avoid the stabilization problems encountered previously with the 
free solid-vapor interface \cite{franck1}.
The $(xy)$ simulation box size was of $22 \times 23 $ 
$\sigma$ units and  
a full fcc layer contained $N_l=480$ atoms. We focused on two 
temperatures, $t_1= 0.75$  and $t_2 = 0.86$, 
(respectively below and above the RL temperature 
 $ t \simeq 0.83$), where we obtained full and converged 
adsorption isotherms. For each temperature 
we increased the chemical potential 
$\mu$ ({\it i.e.}, increased the
pressure of the fictitious perfect gas in contact with the system) by 
intervals of $\simeq\rm 0.02 \epsilon$ and waited for stabilization of both
total energy and particle number. Generally half a
million Monte-Carlo (MC) moves/particle were sufficient to 
reach equilibrium. Then $30$ to $50$ uncorrelated 
configurations were generated 
from a subsequent half million MC moves and analyzed.

Fig. 1 shows the calculated adsorption isotherms -- 
the number of adsorbed layers versus 
 $(\mu_0 -\mu)^{-1/3}$ -- $\mu_0$ being the saturation chemical potential 
(where a bulk quantity of matter would condense). 
At the lower temperature $t_1$ we find clear layering steps 
between consecutive integer layers numbers.
Analysis of layer occupancies shows that after each coverage jump 
the first layer is nearly full, with $\simeq 15-20 \%$
of vacancies, and only few adatoms. In the subsequent plateau the 
adatom population gradually increases to $\simeq 15-20 \%$ and 
vacancies in the first layer are filled, until the next jump
suddenly occurs, and so on.
Between $t_1$ and $t_2$ we generally observed that, as in experiments, 
the layering steps tended to disappear; however here it became 
very difficult to obtain a stable surface and thus well defined
adsorption isotherms.
At the higher temperature $t_2$ we did recover stability,
and we found that layering was again present, but with two important 
qualitative differences with the low temperature isotherm: coverage
was shifted by half a monolayer, and plateaus were broader. 
Adsorption began at  a half-full layer here, and it progressed
continuously, leading to a broader plateau, until the next
jump to another half integer coverage. We plot in 
Fig. 1 
the $t_2$ isotherm up to eight adsorbed layers, 
the maximum thickness before encountering again stabilization problems. 
The large plateau breadths are clearly due to our strong
substrate potential.
The film grand potential can be crudely modeled as a periodic part,
say $k\cos (2\pi n)$, plus an effective interface repulsion, 
$ c/(2 (n-n_0)^2)$ \cite{weeks}, 
plus a growth term $\mu n$ ($n$ is the total number of layers).
The plateau breadth is thus predicted to
decrease asymptotically in the form 
$\Delta n \simeq 1/(1+ \gamma^{-1}(n-n_0)^{4})$, 
where $\gamma=3c/(4\pi ^2 k)$ measures the strength of the substrate. 
As Fig. 1 (inset) shows, this law fits well the experimental
data, with $c/k = 1200$. It also agrees fairly well with our actual
GCMC plateau widths, once $\gamma$ is increased by the correct factor $10$.
We also note from Fig. 1 the relatively large
compressibility $k^{-1}$ of the half-coverage state 
with respect to the low-temperature state.

\begin{figure}
\centerline{\epsfig{figure=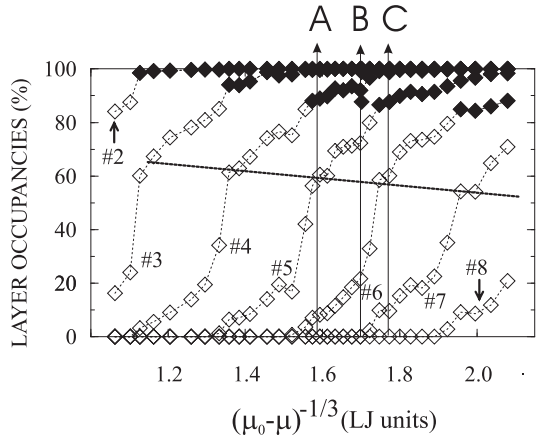,clip=,width=7cm,angle=0}}
\caption{  Occupancies of the different layers versus the chemical
potential calculated at $t_2=0.86$ $(0.61\epsilon)$.
White diamonds indicate a liquid-like layer; black diamonds a solid-like layer.
}
\end{figure}

We conclude that our simulation reproduces the basic RL phenomenon,
making it possible to probe deeply into its nature.
For a better understanding of the layering
reentrance, we plot
in Fig. 2  the occupancies, at $t_2=0.86$, of the different
layers for increasing chemical potential. 
The jumps leading to fractional coverage states 
are clearly observable. Following each jump (A, layer \#5), 
the coverage increases continuously
by a fraction of monolayer, enriching the adatom population, as well as 
first and second layers, until at (B) the surface (layer \#6) is ready for the
next jump, leading to (C) where, following the jump, 
former adatoms (layer \#6) increase in density to form a new half layer, 
and a new adatom layer (\#7) is started. We found no trace of the 
non-monotonic occupancies reported in earlier canonical studies\cite{phillips}.
The top layer occupancy extrapolates to about $50$ \% for large 
adsorbate thickness, strongly supporting the identification with a
DOF state: an ordinary 2D liquid should display a much higher average
lateral density. The 
occupancies of the three outermost
layers (0.1, 0.5, 0.8) for what we thus suppose to describe a 
realistic DOF state differ somewhat from the simplistic ones expected 
from lattice models, namely (0.0, 0.5, 1.0).
The finding of a DOF surface at $t_2$, against an ordinary flat 
surface [occupancies (0.15, 0.85, 1.0)] at $t_1$ indicates that PR of the free
rare gas solid surface must take place in between. This conclusion is also
supported by the evidence of DOF phase separation taking place at 
$t \simeq 0.83$ independently obtained by canonical 
simulations of the free Lennard-Jones surface \cite{jayanthi}.
\begin{figure}
\centerline{\epsfig{figure=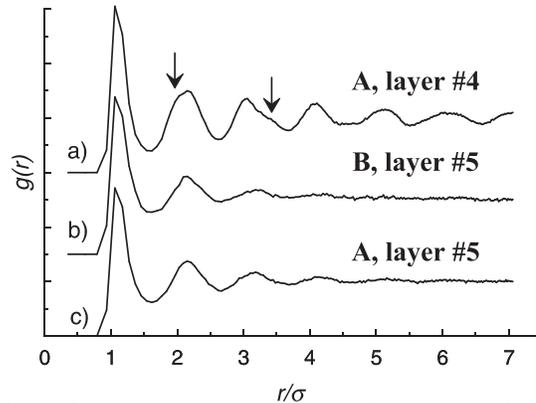,clip=,width=7cm}}
\caption{Pair correlation functions of the states described in the
text. a) Arrows indicate shell features of a solid layer; b),c). 
Absence of shell features indicates liquid layers.
}
\end{figure}

One might thus be led to think that apart from details, the physics
is just that dictated by simple SOS models \cite{weichman}. However, a 
closer look at our MC configurations reveals that the situation is
different, and richer. Following \cite{phillips} we 
studied the lateral positional ordering and diffusion coefficients
of different layers at the 
two temperatures by examining pair correlation functions, in particular 
at $t_2=0.86$. For this purpose we carried out two separate 
canonical molecular dynamics
simulations (the diffusion coefficient is ill-defined in a grand
canonical simulation), one with an integer layer number, and another
with half-integer (no substrate). They were meant to approximate free
stable grand canonical surfaces below and above $t_{RL}= 0.83$, and thus 
chosen with the same coverages $\sim 0.5$ and $\sim 1$ of the 
grand canonical states A and B described earlier.
Fig. 3 shows a selection of lateral pair correlation functions 
$g(r)$ calculated at $t_2$. Presence of shell-related
peaks/shoulders indicates a solid layer, their absence a liquid 
layer\cite{phillips}.
We see that the top layer is always liquid, but that it 
solidifies right after being covered by the next half layer.
Consider for instance state A in Fig. 2.
The upper layer (\#6) has 10\% of adatoms (a 2D gas), the 
lower layer (\#4) has 20\% of vacancies and is solid, 
but the the middle half filled layer (\#5) is liquid.
As coverage increases, layer \#5 gets denser, but  
remains liquid until jump B (see Fig. 3). 
After that, at C, the former adatoms
condense into another fluid half layer \#6, while 
at the same time layer \#5
solidifies, leading to a surface identical to 
the starting one except for one extra layer. This picture is close
to that suggested by heat capacity studies \cite{day}. It is also similar
to that described by canonical simulations \cite{phillips}, differing however
in two crucial respects, namely (i) the lack of a solid-fluid-solid 
evolution for any layer and, more importantly, (ii) the {\em half
occupancy} of the fluid layer. The latter is the hallmark
of the DOF state, which here therefore emerges as the likeliest explanation
for RL. 

In order to further elucidate the connection between surface 
melting and PR, we examined the lateral diffusion coefficient layer by layer.
Mean square displacements were averaged for
all particles spending time within three vertical windows corresponding
to adatom layer, first layer (surface) and second layer. Not surprisingly,  
adatoms are very diffusive (gas-like), while buried layers
are solid, and poorly diffusive. The top layer diffusivity
was always sizable, but larger by about a factor 
two in the half-covered case, where it is similar to the surface mass transport
coefficient near $T_m$ \cite{brough} (Fig. 4). 
This confirms that a height jump by about half a layer across
PR also takes the top layer from solid to liquid, in agreement with
the GCMC analysis.  Thus sudden formation of the liquid half layer at PR
represents the threshold for the first appearance of the liquid, 
which will subsequently extend to lower layers and grow critically
to a thicker liquid film as temperature is further raised to approach 
$T_m$.

\begin{figure}
\centerline{\epsfig{figure=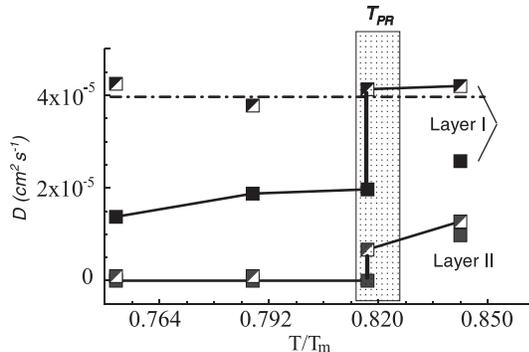,clip=,width=7cm}}
\caption{  
Lateral diffusion coefficient obtained for a free Ar(111) surface with
half (black and white squares) and full (black squares) coverages at different temperatures. The RL onset temperature 
region $T_{PR}$ is dotted. 
At half coverage, the top layer is always liquid; at full coverage, 
the top layer is still solid at $T_{PR}$. The second layer is
always weakly diffusive, and adatoms (not shown) diffuse as a gas 
in every case.
Solid lines guide the eye from full coverage below $T_{PR}$
to half coverage above $T_{PR}$. Note the diffusion coefficient
jump. Surface mass transport coefficient near $T_m$ (from \protect\cite{brough}) is dash-dotted.}
\end{figure}
Summarizing, our results can explain the experimental evidence of RL   
occurring in the adsorption of rare gas on a solid substrate. 
Layer by layer occupancies and direct insight on the surface processes
that are not directly accessible from experiments, 
confirm the interpretation of the reentrant layering transition 
in terms of preroughening. The DOF state consists of a half monolayer of barely percolating 
2D liquid islands, floating on top of a solid substrate.
We found a  coincidence of the onset of premelting in
the top layer with a PR transition, where coverage jumps from full
to partial. These two surface phenomena, apparently very different, 
appear here to be intimately connected. A lattice model addressing 
this connection has been published separately \cite{jagla}.
  
\begin{sloppypar}
It is a pleasure to thank S. Prestipino, E. Jagla, and G. Santoro 
for many constructive discussions. We acknowledge support from INFM, and 
from MURST. Work at SISSA by  F.\ C.\ was under European 
Commission sponsorship, contract ERBCHBGCT940636.
\end{sloppypar}


\begin{references}

\bibitem{SM} see, {\it e.g.}, A. C. Levi, in 
{\it Phase Transitions and Surface Films 2}, Eds H. Taub, G. Torzo, 
H. J. Lauter and S. C. Fain, Jr. (NATO ASI series, Series B, Physics, v. 267), 
p. 327; L. Pietronero and E. Tosatti, Solid State Comm. {\bf 32}, 255 (1979).

\bibitem{vanbejeren} H. Van Bejieren and I. Nolden, in 
{\it Structure and Dynamics of Surfaces II}, Eds. W. Schommers and 
P. von Blanckenhagen (Springer-Verlag, Heidelberg, 1987) p. 259.

\bibitem{dennijs} K. Rommelse and M. den Nijs, 
Phys. Rev. Lett. {\bf 59}, 2578 (1987);
M. den Nijs and K. Rommelse, Phys. Rev. B {\bf 40}, 4709 (1989).


\bibitem{younbook} G. B. Hess, in 
{\it Phase Transitions and Surface Films 2}, {\it op. cit.}, p. 357.

\bibitem{younhess} H. S. Youn and G. B. Hess, Phys. Rev. Lett. 
{\bf 64}, 918 (1990), H. S. Youn, X. F. Meng and G. B. Hess, 
Phys. Rev. B {\bf 48}, 14556 (1993); G. B. Hess, in {\it Phase Transitions 
in Surface Films 2}, edited by H. Taub {\it et al.} (Plenum, New York, 1992).

\bibitem{day} P. Day, M. Lysek, M. LaMadrid, and D. Goodstein,
Phys. Rev. {\bf 47}, 10716 (1993).

\bibitem{comment}J. M. Phillips and J. Z. Larese, Phys. Rev. Lett. {\bf 75}, 4330 (1995); P. B. Weichman and D. Goodstein, Phys. Rev. Lett. {\bf 75}, 4331 (1995).

\bibitem{weichman} P. B. Weichman, P. Day and D. Goodstein, Phys. Rev. Lett. 
{\bf 74}, 418 (1995).

\bibitem{phillips} J. M. Phillips, Q. M. Zhang and J. Z. Larese,
Phys. Rev. Lett. {\bf 71}, 2971 (1993); J.M. Phillips and J. Z. Larese, 
Phys. Rev. Lett. {\bf 75}, 4330 (1995); and Phys. Rev. B {\bf 56}, 15938
(1997).

\bibitem{zippering} P. B. Weichman and A. Prasad, 
Phys. Rev. Lett. {\bf 76}, 2322 (1997).

\bibitem{rieutord}  F. Rieutord, R. Simon, R. Conradt, and P. Muller-Buschbaum,
{\it Europhys. Lett.} {\bf 37}, 565 (1997).

\bibitem{rowley} L. A. Rowley, D. Nicholson and N. G. Parsonage,
J. Comput. Phys. {\bf 26}, 66 (1975).

\bibitem{franck1} F. Celestini, D. Passerone, F. Ercolessi and E. Tosatti, 
Surf. Sci. {\bf 402-404}, 886 (1998).

\bibitem{norman} G. E. Norman and V. S. Filipov, High Temp. (USSR) 
{\bf 7}, 216 (1969).

\bibitem{xjc} X. J. Chen, F. Ercolessi, A. C. Levi and E. Tosatti, Surf. Sci {\bf 249}, 237 (1991).

\bibitem{weeks} J. D. Weeks, Phys. Rev. B {\bf 26}, 3998 (1982).

\bibitem{jayanthi} S. Prestipino, C. S. Jayanthi, F. Ercolessi, 
and E. Tosatti, Surf. Rev. and Lett. {\bf 4}, 843 (1997); 
C. S. Jayanthi {\it et al.}, Surf. Sci. Lett., accepted. 

\bibitem{brough} J. Q. Broughton and G. H. Gilmer, J. Chem. Phys. {\bf 79}(1983)5119.

\bibitem{jagla}E. A. Jagla, S. Prestipino and E. Tosatti , Phys. Rev. Lett. {\bf 83}, 2753 (1999).

\end{references}
\end{document}